\documentclass[acmsmall]{acmart}

\acmJournal{PACMPL}
\acmYear{2022}
\acmMonth{4}
\acmDOI{}
\acmISBN{}
\acmConference{Research Exam}{April 27 2022}{La Jolla, CA, USA}

\startPage{1}

\setcopyright{none}
\usepackage{booktabs}
\usepackage{subcaption}
\usepackage{color}
\usepackage{xspace}
\usepackage[neveradjust]{paralist}
\usepackage{subcaption}
\usepackage{xfrac}
\usepackage{rotating}
\usepackage{multirow}
\usepackage{wrapfig}
\usepackage[normalem]{ulem}
\usepackage[inference]{semantic}
\usepackage[frozencache]{minted}
\newcommand{\scode}[1]{{\texttt{\small #1}}}

\newcommand{\rustc}{\scode{rustc}\xspace}
\newcommand{\unsafe}{\rcode{unsafe}\xspace}

\newcommand{\rcode}[1]{\mintinline{rust}{#1}}

\begin{document}

\title{\mbox{The Usability of Advanced Type Systems: Rust as a Case Study}}

\author{Kasra Ferdowsi}
\orcid{0000-0003-3924-8137}             %% \orcid is optional
\affiliation{
  \position{PhD Student}
  \department{Computer Science \& Engineering}              %% \department is recommended
  \institution{UC San Diego}            %% \institution is required
  \streetaddress{9500 Gilman Drive}
  \city{San Diego}
  \state{CA}
  \postcode{92093}
  \country{USA}                    %% \country is recommended
}
\email{kferdows@eng.ucsd.edu}          %% \email is recommended

\begin{abstract}
    Advanced type systems that enforce various correctness and safety guarantees---such as linear and ownership types---have a long history in the Programming Languages research community.
    Despite this history, a human-centered evaluation of these type systems and their usability was all but absent, with empirical evaluations limited to testing their expressiveness in programs written by experts, i.e. the creators of the type system.

    In the past few years, this has begun to change with the adoption of a version of affine types and ownership in the popular Rust programming language.
    With the increase in Rust's popularity, various studies have begun empirically evaluating the usability of Rust's Ownership and Lifetime rules, providing a breadth of qualitative and quantitative information on the usability of such type systems.
    They found that despite Rust's general success in achieving its promise of safety and performance, these rules come with a steep learning curve and have been repeatedly cited as a barrier to adopting Rust.

    In this report, I provide a brief history of linear types and region-based memory management, which directly inspired Rust's type system.
    I then introduce Rust's Ownership and Lifetime rules, and present the state-of-the-art in academic research into their usability.
    I discuss both theoretical arguments and empirical evidence for why these rules are difficult to learn and apply, and survey existing work on addressing some of these difficulties.
    I also draw from broader works in the HCI and CS Education communities to recommend future work in this area.
\end{abstract}

%% 2012 ACM Computing Classification System (CSS) concepts
%% Generate at 'http://dl.acm.org/ccs/ccs.cfm'.
\begin{CCSXML}
<ccs2012>
<concept>
<concept_id>10011007.10011006.10011008</concept_id>
<concept_desc>Software and its engineering~General programming languages</concept_desc>
<concept_significance>500</concept_significance>
</concept>
<concept>
<concept_id>10003456.10003457.10003521.10003525</concept_id>
<concept_desc>Social and professional topics~History of programming languages</concept_desc>
<concept_significance>300</concept_significance>
</concept>
</ccs2012>
\end{CCSXML}

\ccsdesc[500]{Software and its engineering~General programming languages}
\ccsdesc[300]{Social and professional topics~History of programming languages}
%% End of generated code

%% Keywords
%% comma separated list
% \keywords{Rust, Type Systems, Research Exam}  %% \keywords are mandatory in final camera-ready submission

\pdfinfo{
    /Title (The Usability of Advanced Type Systems: Rust as a Case Study)
    /Author (Kasra Ferdowsi)
}

\maketitle

\section{Introduction}\label{sec:intro}
Despite a plethora of work on advanced type systems in the Programming Languages research community, from Dependent types \cite{xi1999dependent} to Linear \cite{wadler1990linear} and Ownership types \cite{clarke2013ownership}, such type systems have rarely crossed the academic boundaries into mainstream general-purpose programming languages.

A side-effect of this, perhaps exacerbated by a disinterest in human-centered methods in the Programming Languages community in the past \cite{coblenz2018interdisciplinary}, is the lack of any notable user evaluation of such type systems.
So while their usability was repeatedly discussed, the focus was on whether any given type system is ``expressive'', i.e. can an expert write complex and useful code that type checks in that system.
This meant that, until very recently, we simply did not know if such type systems are easy to learn, or how non-experts would learn and use them.

This has begun to change with the Rust programming language \cite{therustlanguage,thebook}.
Rust implement a notion of ``Ownership'', ``Borrowing'', and ``Lifetimes'' as a type system, which allows it to promise memory and thread safety at compile-time without garbage collection.
And with its emergence as an increasingly popular general-purpose programming language, there has come a new wave of research into the usability of its Ownership model.

In this report, I aim to use this new research to better understand the usability of advanced type systems, and to see if and how they may be adopted by mainstream software engineers.
The rest of the report is organized as follows:
\autoref{sec:background} provides a brief background in the history of the type systems most relevant to Rust.
\autoref{sec:rust} then introduces Rust's specific implementation of those type systems, and \autoref{sec:usability} surveys the work on evaluating and improving the usability of Ownership in Rust.
Finally, \autoref{sec:future} combines takeaways from those works with theories from the Human-Computer Interactions and Computer Science Education communities to discuss possible next-steps in research on the usability of advanced type systems.

\section{Background}\label{sec:background}
This section provides a brief overview of Linear Types, Region-Based Memory Management (RBMM) and Ownership types, followed by a higher-level discussion of common themes among them.
This discussion is not meant to be exhaustive, as each system has a long history and would require a separate survey paper by itself\footnote{Which actually exists in the case of Ownership Types \cite{clarke2013ownership}}.
Instead, it is meant to provide a background for the theories that inspired Rust, and to help better connect the usability findings for Rust to type systems more broadly.

\subsection{Linear Types}\label{subsec:linear}
Inspired by Girard's Linear Logic \cite{girard1996linear}, Linear Types were introduced by \cite{wadler1990linear} as a way to safely ``change the world'' (i.e. modify state) in functional programming languages.
The core of Linear Types are values that must be ``used exactly once'', i.e. they cannot be duplicated or implicitly discarded.
Wadler pointed out that this restriction enables a number of static checks and features that may be very useful for memory management and program correctness.

More specifically, he noted that Linear Types enable memory management of mutable values without Garbage Collection.
If a value cannot be copied or implicitly discarded and it must be used exactly once, then we can reclaim its memory after it is used.
This handles memory management, and prevents use-after-free and double-free bugs.
By prohibiting aliasing, Linear Types also solve the problem of reasoning about mutations, both in single-threaded code (where aliased references are a notable source of bugs), and in multi-threaded code (where aliasing can lead to race conditions).

In what will become a common theme in these type systems, Wadler also notes that strict Linear Types are a stronger constraint than necessary, and the language he introduces is not strictly Linear.
Instead, it allows multiple ``read accesses'' (immutable references) to a Linear value that cannot be used once there is a ``write'' access (mutable reference).
I will discuss Linear Types' relation to Rust more in \autoref{sec:rust}, but Rust also uses a looser notion than Linearity called \emph{Affine} types.
Rather than being used \emph{exactly} once, a value with an affine type must be used \emph{at most} once, i.e. it can be ignored \cite{atapl}.

While I will not cover them in depth here, it is worth mentioning that Linear Types have been implemented and extended in various works, from their implementation in the imperative programming language Vault \cite{fahndrich2002adoption} to their recent addition to Haskell \cite{bernardy2017linear}.
They have also been an inspiration for the rest of the type systems in this section.
Linearity is cited in both the seminal works on Region-based Memory Management \cite{tofte1997region} and Ownership Types \cite{clarke1998ownership}, and has affected them over time, with \cite{walker2001regions} combining it with Region types, and the notion of Ownership transfer combining linear and non-linear types \cite{clarke2013ownership}.

\subsection{Region-based Memory Management}\label{subsec:region}
As the name suggests, Region-based Memory Management (RBMM) was an effort in static memory management using the type system.
RBMM started as an extension of Effect Type Systems \cite{atapl} in \cite{tofte1997region}.
But its implementation for the Cyclone language \cite{cyclone,jim2002cyclone,grossman2002region} is the more direct influence on Rust, and so I will focus on that here.

Cyclone began as a part of the Typed Assembly Language project \cite{morrisett1999talx86}, but was developed into a separate project aiming to become ``a safe dialect of C'' \cite{grossman2005cyclone}.
As such it contains a number of interesting design choices and language features besides RBMM such as tagged unions, null checks, and existential types.
\cite{grossman2005cyclone} offers a concise description of all these features, but here I will focus on RBMM in particular as described in \cite{cyclone}, as it is the most relevant to this report.

The key idea of RBMM is to associate a lexically-scoped part of the program with a named ``region'' (a dynamically growable part of memory), and annotate the type of pointers into that region with the region's name.
Then, the compiler can automatically deallocate the entire region at the end of the scope, and the type-checker can guarantee that pointers into a region are not dereferrenced outside of that region's scope (i.e. after it is deallocated).
To do this, the type system keeps track of the set of regions that are live at each point in the program (called the ``capability'' at that point), and prohibits pointer dereferencing unless that pointer's associated region is in the capability.

This is the core of RBMM, but to make it sufficiently expressive and guarantee soundness, there's a lot more subtlety involved.
For example, region types can support \emph{subtyping}.
Since regions can be nested, all pointers into the outer region are guaranteed to be alive during the inner one (since the outer region is deallocated after the inner).
So if a region \rcode{'a} contains a smaller region \rcode{'b}, pointers annotated with \rcode{'a} are a subtype of \rcode{'b}.

Another detail that Cyclone developers considered was the syntactic overhead of region annotations, and the need for region generics (functions with arguments and return types that are generic over region annotations).
Their solution was a combination of intraprocedural region annotation inference, which removed the need for most explicit annotations in function bodies, and defaults for partially-annotated function signatures.
This removed a large amount of the syntactic overhead, and made certain functions translate from C to Cyclone directly with no or minimal change.

Outside of Cyclone, RBMM has had a long history. It has been implemented for the Go programming language \cite{davis2012towards}, Real-Time Java \cite{boyapati2003ownership}, Prolog \cite{makholm2000region}, GPU programming \cite{hold2014region}, and Big Data systems \cite{nguyen2015speculative}.
But none of these works empirically evaluated the usability of their system on programmers, focusing instead on benchmark performance and limiting their discussion of usability to expressiveness and syntactic overhead.

\subsection{Ownership Types}\label{subsec:ownershiptypes}
Despite having a similar name, Ownership Types are not directly related to Rust's notion of Ownership.
However, they share many of their goals with Rust, and are a key part of the history of Ownership as a concept.
So a discussion of relevant type systems for Rust would be incomplete without them.

Ownership Types were developed as a part of Object-Oriented Programming (OOP) to statically enforce a more strict notion of ``encapsulation'' \cite{clarke2013ownership}.
While the details vary greatly between implementations, the general idea is to encode an ``owning'' and ``owner'' relationship between objects in the type system. This places restrictions on pointer aliasing which enforce encapsulation \cite{clarke1998ownership}, and enable additional checks and guarantees, such as preventing data races and deadlocks \cite{boyapati2002ownership} and dangling pointers \cite{boyapati2003ownership}.

Despite the large body of work on Ownership types, and its close relation to Java (a popular general-purpose programming language), these types were neither widely adopted, nor evaluated on real users.
Instead, each extension, implementation or application of these types was only evaluated by the designers of the system, who programmed real-world applications with their type system to argue for its expressiveness \cite{clarke1998ownership,clarke2013ownership,boyapati2003ownership,boyapati2003ownershipagain,aldrich2002alias}.

Despite having a mostly separate history, Ownership Types are conceptually very close to the other type systems in this report. For instance, in Ownership types ``a program's heap is divided into hierarchically nested regions, originally called ownership contexts'' \cite{clarke2013ownership} which is similar to regions in RBMM.
In fact, \cite{boyapati2003ownership} combined Ownership types with Region-based Memory Management to implement a type system for Real-Time Specification for Java (RTSJ).
This system statically guaranteed the success of runtime checks for dangling pointers, and a lack of references to the garbage-collected heap (a requirement in RTSJ code).
However, similar to the rest of the works in this section, this system was never evaluated on real users.

\subsection{Shared Themes}\label{subsec:themes}
So far I introduced each type system individually, but these ideas and systems are closely related, and their development is not easily separable.
So, before moving to Rust, I will first discuss the broad trends in these works more holistically.

Linear, Region and Ownership Types are related by their attention to memory management and safety.
Each realized that type systems could be used to help programmers reason about complex programs,
prevent various errors in using aliased or freed references,
and offer a provably correct solution to memory management without the need for runtime checks or garbage collection.

They were all also quick to note and try to tackle the trade-off between the ``expressiveness'' of their type systems, and their guarantees.
In the paper that introduced Linear Types, \cite{wadler1990linear} did not enforce Linearity, but allowed combining values of Linear and non-Linear types, and (as discussed above) loosened the definition of uniqueness to allow multiple read-only references to Linear values.
Despite touting RBMM, Cyclone \cite{grossman2002region} included a distinguished garbage-collected heap region.
And a few years after its introduction, early proponents of Ownership Types were already making the case that strict uniqueness is needlessly restrictive \cite{clarke2003external}.

Finally, despite this attention to expressiveness, a tendency to implement their type systems as versions or extensions of popular programming languages (ML \cite{tofte1997region}, C \cite{grossman2002region}, Java \cite{aldrich2002alias}, Scala \cite{haller2010capabilities}, Go \cite{davis2012towards}, Prolog \cite{makholm2000region}, etc.), and even considering the benefit of such restrictions in program comprehension \cite{aldrich2002alias,clarke2013ownership}, a user-centered approach was missing from all the works cited above.
No one studied if users other than those who had invented and implemented the type systems could easily work with the restrictions imposed by them.

\section{The Rust Programming Language}\label{sec:rust}
Rust, though inspired by \cite{wadler1990linear} for its notion of Ownership, and \cite{grossman2002region} for its approach to lifetimes and memory management, does not directly implement any of the type systems above.
Instead, it combines them with a number of other ideas to try to guarantee memory- and thread-safety, as well as static memory management without garbage collection.
However, it also aims to be a general-purpose systems programming language\footnote{The term ``Systems Programming Language'' has caused some controversy recently \cite{crichton2018what}. But given its colloquial use as a language that compiles to assembly, and offers low-level control of resources (as opposed to interpreted languages like Python, or those that run on higher levels of abstraction such as Java), I will use that term in this report for simplicity's sake.},
and so aims for ``pragmatic safety'' \cite{evans2020is} and has features that bypass its static checks for better performance or more complex aliasing patterns.

In the rest of this section, I will first introduce Ownership as it is used in Rust, then describe lifetimes and how they combine with Ownership, and finally describe unsafe code which bypasses some of these checks.
I have restricted my descriptions here to what's necessary to think about some of the usability issues I will discuss in \autoref{sec:usability}, and I am ignoring many subtleties of the type system, as well as any mention of Rust's syntax, semantics, etc.
For a good introduction to Rust in general, I recommend the official Rust book \cite{thebook}, which is both thorough and very readable.

\subsection{Ownership}\label{subsec:ownership}

\begin{figure}
    \begin{subfigure}[t]{0.3\textwidth}
        \begin{minted}[fontsize=\footnotesize,linenos,autogobble]{rust}
          let v = vec![1, 2];
          let v2 = v;
          print(&v);
      \end{minted}
      \caption{}
      \label{fig:rules:1}
  \end{subfigure}
  \begin{subfigure}[t]{0.3\textwidth}
      \begin{minted}[fontsize=\footnotesize,linenos,autogobble]{rust}
          let v = vec![1, 2];
          let x = &v[0];
          let v2 = v;
          let y = *x + 1;
      \end{minted}
      \caption{}
      \label{fig:rules:2}
  \end{subfigure}
  \begin{subfigure}[t]{0.3\textwidth}
      \begin{minted}[fontsize=\footnotesize,linenos,autogobble]{rust}
          let mut v = vec![1, 2];
          let x = &v[0];
          Vec::push(&mut v, 3);
          let y = *x + 1;
      \end{minted}
      \caption{}
      \label{fig:rules:3}
  \end{subfigure}
  \caption{Each subfigure demonstrates a violation of the corresponding rule in \autoref{subsec:ownership}. In \autoref{fig:rules:1} ownership of the vector is transferred to \rcode{v2} on line 2, so \rcode{v} cannot be borrowed for the call to \rcode{print} on line 3.
In \autoref{fig:rules:2} \rcode{x} is a reference to \rcode{v} and lives until its last use on line 4. But \rcode{v} only lives until the transfer of its vector to \rcode{v2} on line 3.
Similarly, in \autoref{fig:rules:3}, the \emph{immutable} borrow of \rcode{v} in \rcode{x} lives from line 2 to its last use on line 4, so \rcode{v} cannot be \emph{mutably} borrowed by the call to \rcode{push} on line 3.}
  \label{fig:rules}
\end{figure}

Ownership rules in Rust are, on paper, quite simple, and various papers have attempted to summarize them \cite{qin2020understanding,fulton2021benefits}.
Here I will use \cite{crichton2020usability}, since it is the most simple and concise.
But first, I need to introduce some terms.

In Rust, each value (a \rcode{String}, \rcode{i32}, \rcode{Vec}, etc.) is owned by a single variable, which is its \textbf{owner}.
Since working with values directly can be inconvenient, Rust also has \textbf{references} to values which \textbf{borrow} the value from its owner.
Finally, in Rust variables and references are \textbf{immutable} by default.
To mutate a value through one, it needs to be explicitly marked as \textbf{mutable} using the \rcode{mut} keyword.
For instance, in \autoref{fig:rules:3} line 1, the \rcode{Vec} created by the call to \rcode{vec!} is assigned to the variable \rcode{v}, thus \rcode{v} owns that \rcode{Vec}.
On line 2, \rcode{x} is a reference to the first element of \rcode{v}, and thus \rcode{x} borrows \rcode{v}. Finally, \rcode{v} is marked with the \rcode{mut} keyword, and thus it is mutable, but \rcode{x} is not, and so \rcode{x} can only be read, not modified or reassigned.

With ownership, references and mutability in mind, \cite{crichton2020usability} summarizes Rust's Ownership rules like so:
\begin{enumerate}[(a)]
\item All values have exactly one owner.
\item A reference to a value cannot outlive the owner.
\item A value can have one mutable reference or many immutable references.
\end{enumerate}
See \autoref{fig:rules} for a code example of what violating each of these rules may look like.
Rust's compiler \scode{rustc} contains a pass, known colloquially as the borrow-checker, which fails if it cannot statically determine that all of these rules are being followed.

Before moving on to lifetimes, it is worth considering how these rules relate to the type systems in \autoref{sec:background}.
Rather unintuitively, Rust's Ownership rules are not directly related to Ownership Types discussed in \autoref{subsec:ownershiptypes}\footnote{This confusion is further exacerbated by more recent papers such as \cite{crichton2021modular} which refer to Rust's Ownership rules as ``Ownership Types'', despite the existing history of Ownership Types as a different system.}.
Rather, the first rule which is generally referred to as ``Ownership'' is most closely tied to Linear Types: A value has a single owner at any point in the program, and while Ownership can be transferred between variables, it cannot be implicitly duplicated.

This is not to say that Ownership Types are entirely unrelated.
Mutability and borrowing are more explicitly dealt with in Ownership Types than Linear or Region types.
Certain instances of Ownership Types restrict mutation to the owner of a value, and only permit read-access from other objects.
Similarly, the notion of borrowing appears in parts of the Ownership Type literature with a similar function \cite{boyland2001alias,aldrich2002alias}.
But Ownership Types are closely tied to concepts from Object-Oriented Programming (which is not Rust's paradigm). And I have only found a single mention of Ownership Types as an influence on Rust in the literature \cite{weiss2021oxide}.

I will leave a detailed comparison to RBMM to the next section, but note how rules (a) and (b) also allow automatic memory management:
When the owner of a value goes out of scope, it is guaranteed not to have any live references, and so the compiler can insert a call to deallocate that value (\rcode{drop} the value in Rust parlance).
This preserves memory-safety without the need for garbage collection.

\subsection{Lifetimes}\label{subsec:lifetimes}

\begin{figure}
    \begin{subfigure}[t]{0.45\textwidth}
    \begin{minted}[fontsize=\footnotesize,linenos,autogobble]{rust}
        struct Foo<'small, 'large: 'small> {
            a: &'small str,
            b: &'large str,
        }
        impl<'s, 'l: 's> Foo<'s, 'l> {
            fn new(x: &'s str, y: &'l str) -> Self {
                Self { a: x, b: y }
            }
        }
    \end{minted}
\end{subfigure}
\hspace{0.05\textwidth}
    \begin{subfigure}[t]{0.40\textwidth}
    \begin{minted}[fontsize=\footnotesize,linenos,autogobble,firstnumber=last]{rust}
        fn main() {
            let mystr = "abc";
            let substr = &mystr[0..2];
            let foo = Foo::new(&mystr, substr);
        }
    \end{minted}
    \end{subfigure}
    \caption{An example of explicit lifetime parameters in \rcode{struct} and \rcode{fn} definitions. The lifetime parameters define the \rcode{struct} and \rcode{fn} as generic over lifetimes \rcode{'small} and \rcode{'large}, where \rcode{'large} is a subtype of \rcode{'small}. Rust then uses these parameters to type check the use of \rcode{Foo::new} in the \rcode{main} function by comparing the inferred lifetimes for the references passed to \rcode{Foo::new} with the explicit lifetime parameters. This code compiles because \rcode{substr} borrows \rcode{mystr} and so its lifetime must be smaller than \rcode{mystr}, which satisfies the subtyping requirement between \rcode{'small} and \rcode{'large}.}
    \label{fig:lifetimes}
\end{figure}

\cite{fulton2021benefits} gives a great and concise description of lifetimes:
\begin{quote}
``A lifetime names a scope, and a lifetime annotation on a reference tells the compiler the reference is valid only within that scope.''
\end{quote}
Lifetime annotation here refers to the fact that Rust references are not the same as C-style ``raw'' pointers.
A raw pointer's type only has the type of the value it points to.
But the type of a Rust reference is annotated with a lifetime that refers to the scope where that reference is valid, and lets the borrow-checker keep track of which value (or other reference) it is borrowing.

Rust automatically infers all lifetimes in function bodies, and so most annotations are not visible to the user.
However, when writing functions that have references in their signatures, or data types which store references, Rust requires users to explicitly write them as generic functions/data types over the lifetimes of those references\footnote{There is an exception to this, which is functions whose signatures follow a particular pattern such as functions that don't return a reference or which take exactly one reference and return a reference. In these cases, Rust \emph{elides} these lifetimes as a syntactic convenience. Note that this is not the same as lifetime inference. Rust, similar to Cyclone, does \emph{not} infer lifetimes in function signatures.}.
You can see examples of this in \autoref{fig:lifetimes}.

There is much more to lifetimes, how they are calculated, and their implications on expressiveness and usability. But, as we shall see, Rust lifetimes are notoriously complex and difficult, and a full description of these aspects is outside the scope of this report.
So I will leave lifetimes here, and end this section with a discussion of their relation to Cyclone.

As the reader may have noticed by now, lifetimes in Rust are very similar to regions and region annotations in Cyclone, including their syntax, subtyping, and intraprocedural inference.
In fact, many of Rust's features were inspired by Cyclone, such automatic bounds checking, and sum types (Rust \rcode{enum}s and Cyclone's tagged unions).

The main difference between the two is that Cyclone requires manually defining the syntactic scope of regions, and using the region names (which are first-class values) to manually allocate and initialize values inside different regions.
In Rust, lifetimes are automatically determined by the compiler, and cannot be explicitly set or used (except for generic lifetime parameters).
Also, since Rust 2018, Rust regions are not determined lexically, but are instead calculated over an intermediate control-flow graph representation of the program \cite{rust2018}.
So while the theory behind regions and lifetimes is the same, their implementation and interaction model are different in interesting and significant ways.

\subsection{Unsafe Rust}\label{subsec:unsafe}
One of the common themes among the type systems discussed in \autoref{sec:background} was that each found strict adherence to its rules needlessly restrictive, and found ways to loosen it for the sake of expressiveness.
Rust is no exception to this, though its solution is rather different.

An important issue with Rust's Ownership rules is that they are sound, but undecidable.
So the borrow-checker is incomplete, and there is plenty of safe code which follows the Ownership rules, but the borrow-checker cannot statically verify.
To alleviate this, Rust allows users to explicitly mark functions and blocks of code as \unsafe, and in these unsafe blocks, certain safety checks are disabled\footnote{Some, \cite{zhu2022learning} for example, refer to unsafe code as ``similar to the C Programming Language''.
While this is technically true because unsafe code can interface with arbitrary C code, unsafe code within Rust is still far more restrictive than C.}.
More specifically, \unsafe allows the code to:
\begin{itemize}
\item Dereference raw pointers
\item Call unsafe functions (including C functions, compiler intrinsics, and the raw allocator)
\item Implement unsafe traits
\item Mutate statics
\item Access fields of unions
\end{itemize}
\cite{rustonomicon}.
This is still quite restrictive, but has serious implications.
For instance, dereferencing raw pointers can work around the Ownership rules by casting a reference with one lifetime into a raw pointer, and dereferencing that raw pointer to borrow it again as a new reference \emph{with a new lifetime}. This allows creating multiple mutable references to a value at the same time, which violates the third rule of Ownership.

The details of this are again notoriously complex and beyond the scope of this report, but unsafe code is crucial to Rust's ``pragmatic safety''.
It allows various performance improvements that are too low-level for the borrow-checker to reason about, as well as interfacing with external code, and certain aliasing patterns which the programmer can verify as safe, but do not pass the borrow-checker.

\section{The Usability of Ownership}\label{sec:usability}

At the time of writing, there have been five major papers on the usability of Rust's Ownership type system \cite{zeng2019identifying,fulton2021benefits,bronzegc,crichton2020usability,zhu2022learning}.
And, following the best practices of behavioral research \cite{mcgrath1995methodology}, they use a variety of methods to inspect a number of similar and overlapping research questions.
So, rather than review each paper individually, in this section I discuss their collective findings, introducing the papers and their methodology as they become relevant.
I also draw from research on the use of \unsafe in Rust, as well as related research on how programmers learn a new programming language.

The high-level takeaway is that Rust's Ownership model is indeed difficult to learn, and certain aspects of its design remain difficult even for more experienced Rust developers.
However, its promise of safety and performance, coupled with good tooling and features for interoperability with other languages, keep Rust popular and loved by those who succeed in adopting it.

\subsection{Is Ownership difficult to use?}\label{subsec:usability:1}
\begin{quotation}
    ``Learning Rust Ownership is like navigating a maze where the walls are made of asbestos and frustration, and the maze has no exit, and every time you hit a dead end you get an aneurysm and die.''

    \hspace{1em plus 1fill} --- Student participant from \cite{bronzegc}
\end{quotation}

Rust is notorious for it's ``steep learning curve'', and this has been noted as a major issue in adopting it in the industry. But interestingly, studies suggest that even experienced developers struggle with certain aspects of Ownership in Rust.

\subsubsection{Barriers for Novice Rust Programmers}
\cite{zeng2019identifying} performed content analysis on top posts from the \scode{/r/rust} subreddit (an online community specific to Rust), and articles and corresponding comments from Hacker News (a broader tech forum) to identify barriers to adopting Rust.
In 18 experience reports and language comparisons they inspected, they found that ``the complexity of the borrow-checker was the second most frequently mentioned complaint'' (second only to compiler version issues), where memory access patterns that were common in other languages were disallowed by the borrow-checker, leading to frustration.

\cite{fulton2021benefits} followed this work by interviewing 16 industry professionals who had attempted to adopt Rust in their production team, and used their findings to design an online survey which provided them with 178 more participants. They also found that Rust's steep learning curve was the most serious barrier to adoption.

Note that this difficulty is more than simply the difficulty of learning a new language, or indeed learning a systems programming language without garbage collection.
\cite{fulton2021benefits} found that the biggest challenge in learning Rust was specifically the borrow-checker, and the necessary shift in programming paradigm to write code that passes the borrow-checker.
And \cite{shrestha2020here} quoted a C++ developer who said that the borrow-checker ``forces a programmer to think differently''.
So it appears that Rust's more advanced type system is the main source of its difficulty, not just a lack of garbage collection, or more low-level programming.

Which is not to say that adding garbage collection will not ease Rust's difficulty.
\cite{bronzegc} ran a controlled study on 428 students in a sophomore-level programming course.
The students were given two weeks of lectures on Rust, and then asked to complete an assignment which required a good understanding of Rust, its Ownership rules, and types that allow for interior mutability.
They randomly assigned students to two groups, one having to complete the assignment using the Rust standard library data types, and one using a garbage-collected wrapper type (called ``Bronze'') which enabled a number of additional aliasing patterns to pass the borrow-checker, thus removing the needs for more complex aliasing patterns and datatypes.

They found a significant difference in the rate of completion and the self-reported time to completing the assignment.
The students who used Bronze on average took only a third as much time as the control group, and were approximately 2.44 times more likely to complete the assignment.
Interestingly, the time difference between the groups only appeared in the second part of the task, which involved complex aliasing and mutability.
The first part of the assignment, which focused just on Ownership, didn't show a significant difference between the groups.

\subsubsection{Barriers for Experienced Rust Programmers}
Aside from the initial learning curve, studies also suggest that aspects of Ownership remain difficult to use, even for experienced developers.
In their study of memory- and thread-safety issues in Rust, \cite{qin2020understanding} inspected five Rust systems and applications, five popular libraries, and two vulnerability databases.
They found that a common reason for blocking bugs in these codebases was a lack of ``good understanding in Rust's lifetime rules''.
This is notable since, unlike the participants in the studies above, the programmers who worked in these codebases were presumably experienced Rust developers.

This finding is corroborated by results from the Rust community's 2020 survey \cite{rust2020survey}.
They received 8323 responses, with the largest number of participants self-reporting their expertise as 7 out of 10.
They also found that lifetimes are the most difficult topic to learn, though unfortunately they do not report if and how this response changes according to the expertise rating.
Similarly, in their study of 100 samples of StackOverflow questions on Rust's Ownership rules, \cite{zhu2022learning} found that the most common cause of safety rule violations in these questions was ``complex lifetime computation'', which appeared 74 times\footnote{The paper counts 77 violations, but I'm exlcluding 3 which were merely syntax errors.}, 44 in intraprocedural lifetime computation, 16 in explicit lifetime parameters, and 14 in elided ones.

From these works, it seems safe to conclude that Rust's Ownership rules are indeed difficult to learn. They pose a serious barrier to learning and adopting Rust, and understanding lifetimes specifically remains a problem even for more experienced Rust developers.

\subsection{Why is Ownership difficult to use?}\label{subsec:usability:2}

\begin{quotation}
   ``I can teach the three rules [of Ownership] in a single lecture to a room of undergrads. But the vagaries of the borrow checker still trip me up every time I use Rust!''

    \hspace{1em plus 1fill} --- \cite{crichton2020usability}
\end{quotation}

If we accept that Rust is indeed more difficult to learn than comparable systems programming languages, and that this is in large part caused by its Ownership type system specifically, the next step is to ask what about Rust's Ownership rules is difficult to learn and apply.

\subsubsection{Change of Paradigm}
One answer may be the notion of ``interference'' as used by \cite{shrestha2020here}.
In that paper, they qualitatively coded 450 posts on StackOverflow for 18 different programming languages, and interviewed 16 professional programmers, to understand how experienced developers learn new programming languages, and what they struggle with in the process.
They motivated this work by borrowing the term ``interference''\footnote{As well as the term ``facilitation'', but that is not as relevant here.}
from psychology and neuroscience.
The term refers to when ``previous knowledge disrupts recall of newly learned information''.
This can be as simple as the difference in zero- vs. one-indexing between two languages, but it also applies to larger differences, where programming in the new language requires a ``mindshift'', or a fundamental change in paradigms.

Learning Rust needs such a mindshift, because its Ownership rules prohibit many common programming patterns. Consider a doubly-linked list. In most languages its implementation is close to trivial, but it violates Rust's rules by definition: It requires at least two mutable references to a node, one from the previous and one from the next node. Rust has workarounds for this, most simply datatypes with ``interior mutability'' that postpone checking for simultaneous mutable access to runtime, but they are more difficult to learn and work with. So it is unsurprising that \cite{shrestha2020here} use Rust's Ownership type system as an example of mindshifts, quoting a C\# developer who had to ``completely rethink the problems they would have normally solved in C\#''.

In the qualitative portion of their study, \cite{bronzegc} noted a similar theme in the students' survey responses.
For the students without the Bronze library, the second part of the assignment required using types with interior mutability and explicit lifetime parameter declarations.
Students mentioned the difficulty of using these types, and the need for redesigning their code to use them correctly, leading the authors of the paper to conclude that ``most of the benefit of GC comes from architectural simplification'' and that ``design was a significant contributor to the difference in performance between non-Bronze and Bronze participants.''

So at least one main reason for the difficulty of learning Ownership is that it requires a change of paradigm.
A programmer who is new to Rust needs to learn entirely new patterns and ways of structuring code at the architectural level.
And their previous experience can actively interfere with their learning, as they need to abandon common programming patterns and learn to structure their code in new and unintuitive ways.

\subsubsection{Error Messages}
\cite{bronzegc} also noted that \scode{rustc}'s error messages contributed to the confusion and frustration.
\scode{rustc} error message not only describe the error in the code, but for certain error patterns, suggest edits that may fix the problem.
However, these edits are always local and don't provide any high-level design feedback which may be helpful in making the mindshift.
At best, they led the students to perform a chain of local edits that resulted in code that compiles without them understanding why.
At worst, as one student found, they could be cyclical ``with things like \scode{remove \&} then after removing \scode{try adding \&}.''
This lead the authors to conclude that  Rust's error messages do not ``aid design or comprehension''.

\cite{zhu2022learning} investigated error messages more directly.
They employed Cognitive Task Analysis \cite{diaper2004understanding} to learn how experts solve 110 Rust Ownership errors they had identified in a sample of StackOverflow questions,
and compared the steps the experts took to the information contained in the error message.
They found that while for most errors the error message contained all relevant information, for 32 errors the message failed to explain ``the key steps in computing a lifetime or a borrowing relationship'', with another 10 failing to ``explain the relationship between two lifetime annotations'', and 9 ``how a safety rule works on a particular code construct''.

I will leave a broader discussion of \rustc error messages to \autoref{sec:future}, but the works cited here indicate that Rust error messages do not provide the necessary help.
Programmers' errors may be more structural, but the error messages only suggest potentially misleading local edits.
And even for local errors, they do not always contain the necessary information to understand and fix the error, and assume external knowledge on behalf of the programmer.
But this still doesn't explain why an \emph{experienced} Rust developer struggles with Ownership.

\subsubsection{The Curse of Incompleteness}

\begin{figure}
    \centering
    \begin{subfigure}[t]{0.3\textwidth}
        \begin{minted}[fontsize=\footnotesize,linenos,autogobble]{rust}
          let mut v = vec![1, 2];
          let one = &mut v[0];
          let two = &mut v[1];
          *two += *one;

      \end{minted}
      \caption{}
      \label{fig:incompleteness:1}
  \end{subfigure}
  \begin{subfigure}[t]{0.30\textwidth}
      \begin{minted}[fontsize=\footnotesize,linenos,autogobble]{rust}
          let mut v = vec![1, 2];
          let iter = v.iter_mut();
          let one = iter.next().unwrap();
          let two = iter.next().unwrap();
          *two += *one;
      \end{minted}
      \caption{}
      \label{fig:incompleteness:2}
  \end{subfigure}
  \hspace{0.075\textwidth}
  \begin{subfigure}[t]{0.225\textwidth}
      \begin{minted}[fontsize=\footnotesize,linenos,autogobble]{rust}
          let mut v = vec![1, 2];
          v.insert(0, v[0]);
          v.get_mut(v[0]);


      \end{minted}
      \caption{}
      \label{fig:incompleteness:3}
  \end{subfigure}
  \caption{The programs in \autoref{fig:incompleteness:1} and \autoref{fig:incompleteness:2} perform the same function, but only \autoref{fig:incompleteness:2} passes the borrow-checker. In \autoref{fig:incompleteness:3}, the statements on lines 2 and 3 are nearly identical at the type-level, but only line 2 passes the borrow-checker, presumably due to some implementation detail.}
  \label{fig:incompleteness}
\end{figure}

\cite{crichton2020usability} point out that Ownership rules are simple and easy to learn, but statically checking for them, ``like most interesting program properties'', is undecidable.
So Rust's implementation of these rules in the borrow-checker is necessarily incomplete, and a lot of the usability issues with Ownership come from this gap between the programmer's understanding of the rules, and the borrow-checker's ability of verify them.

Consider the examples in \autoref{fig:incompleteness:1} and \autoref{fig:incompleteness:2}.
Both programs perform a similar function, getting references to two elements of a vector and incrementing one by the other.
But only the second compiles, since the borrow-checker cannot reason about indices.
It conservatively assumes that both lines 2 and 3 in \autoref{fig:incompleteness:1} are mutably borrowing the entire vector, thus violating Rule (c) in \autoref{subsec:ownership}.
\autoref{fig:incompleteness:2}, however, uses an iterator, which the borrow-checker can reason about at the type level\footnote{I should mention that \rcode{iter_mut} uses \unsafe to achieve this under the hood, but since \rcode{iter_mut} is itself a safe function provided by the Rust standard library, it can easily be used by novices without ever touching unsafe code.}, and can successfully verify does not violate the Ownership rules.
Thus \autoref{fig:incompleteness:2} compiles successfully.
Note that both of these programs are ``safe'', and a more advanced type system involving dependent types could in theory statically verify the safety of \autoref{fig:incompleteness:1}, but the current limitations of the type system means that developers need to learn, not just the rules of Ownership, but how the borrow-checker verifies them.

This issue is exacerbated by the fact that the borrow-checker's implementation is quite complex and sometimes very similar code may not compile for obscure reasons.
The code example in \autoref{fig:incompleteness:3} has two similar calls to functions on the vector.
Line 2 gets the first element of \rcode{v}, and inserts it as the new first element of \rcode{v}.
Line 3 uses the \emph{value} of the first element as an index to get a mutable reference to the second element of \rcode{v}.
These functions have very similar types, both using an immutable borrow of \rcode{v} to get an argument for a call that mutably borrows \rcode{v}.
However, as of Rust version \rcode{1.59.0}, line 2 compiles successfully, but line 3 fails the borrow-checker\footnote{\cite{crichton2020usability} speculates that the reason for this is that \rcode{get_mut} is defined on slices (which the \rcode{Vec} type implements), while \rcode{insert} is implemented directly on \rcode{Vec}. They don't know why this distinction matters, and it only further proves their point.}.

There are almost certainly many other large and small, obvious and subtle reasons for the difficulty of learning and using Rust's Ownership type system, but these three (Rust's different paradigm, unhelpful error messages, and the incompleteness of the borrow-checker) are the most apparent from the works surveyed here.

\subsection{Why do developers try to use Rust anyway?}\label{subsec:usability:3}
\begin{quotation}
    Instead of having to invoke \scode{pkg-config} by hand or with Autotools macros, wrangling include paths for header files and library files and basically depending on the user to ensure that the correct versions of libraries are installed, you write a \scode{Cargo.toml} file which lists the names and versions of your dependencies. [...] It just works when you \scode{cargo build}.

    \hspace{1em plus 1fill} --- \cite{federico2018rust}\footnote{Quote found in \cite{zeng2019identifying}.}
\end{quotation}

The last question I will inspect here is that, if Ownership is difficult to learn and use for so many reasons, why do developers choose to use Rust anyway?

And perhaps the first answer to that is that they don't.
Despite being the ``Most Loved'' programming language in every StackOverflow survey since 2016 \cite{so2016survey,so2017survey,so2018survey,so2019survey,so2020survey,so2021survey}, it's user-base is small and growing slowly.
In the same surveys, it appeared in the list of Most Popular languages in 2019 at only 3.2\% \cite{so2019survey}, growing to 7.03\% in the latest survey \cite{so2021survey}.
In comparison, the Go programming language (which is often compared with Rust as a modern systems programming languages) was already at 8.2\% in 2019, though it only grew to 9.55\% by 2021.
Similarly, the TiOBE index ranks Rust at 26 \cite{tiobe}, and the IEEE Spectrum ranks it at 17 \cite{ieeespectrum}, compared to 13 and 8 for Go.
There could be many reasons for this beyond the usability of Ownership of course, and these are not peer-reviewed sources.

But Rust's popularity \emph{is} still growing, and unsurprisingly the main reason most participants in multiple studies cited was its promise of memory- and thread-safety \cite{zeng2019identifying,fulton2021benefits}.
Unfortunately, neither of these papers go into depth about this, and only mention that safety is the most commonly noted reason.
However, other themes besides safety emerged in these works that are far more interesting.

The first is that while safety is important, it is not enough.
\cite{zeng2019identifying} noted that while the first and third most-noted benefits of adopting Rust were avoiding runtime errors and data races, the second most-mentioned benefit was Rust's build tool \scode{cargo}, which avoided the many issues of build tools for other langauges.
Similarly, while \cite{fulton2021benefits}'s participants cited Rust's safety as a benefit most frequently, they listed performance and lack of garbage collection almost as frequently.

Another theme that came up in multiple works was Rust's \unsafe feature.
Two studies which inspected the use of \unsafe in Rust code repositories found that unsafe code is common.
\cite{evans2020is} inspected all publicly available Rust libraries on \scode{crate.io} (Rust's online library registry), and found that explicit \unsafe blocks appear in 29\% of all libraries.
When they filtered their results to the most popular libraries (which accounted for 90\% of downloads), this percentage increased to 52.5\%.
In the 5 applications they inspected, \cite{qin2020understanding} found 4990 uses of \unsafe, with a further 1581 unsafe code regions in the standard library, and concluded that unsafe code is used ``extensively''.
Though they note that it is ``unavoidable in many cases'' and ``usually for good reasons'', including interfacing with existing libraries written in other unsafe languages such as C, and performance improvements by a factor of 4 or 5.
Interestingly, they also found cases of the \unsafe keyword being used as a warning to developers, despite the code itself being safe and compiling without the unsafe block.
Those who tried to adopt Rust also noted the many uses for unsafe code, citing its necessity for integrating Rust into existing codebases through FFIs, accessing hardware, and for performance reasons \cite{fulton2021benefits}.
So it seems that ``pragmatic safety'' was an essential part of Rust's success, as a large amount of code written in Rust would have not been possible it if had strictly adhered to its statically-guaranteed safety rules.

It's also interesting to note that we now have empirical evidence that, despite \unsafe being described as an ``escape hatch'' \cite{evans2020is}, developers can be trusted to use it responsibly, only where necessary and while still mostly maintaining Rust's safety guarantees.
As \cite{evans2020is} found, most Rust codebases do not contain explicit unsafe code.
And in their study of StackOverflow questions, \cite{zhu2022learning} found that of the 110 errors in the questions, only 3 were fixed by writing unsafe code.
The rest were either fixed with simple safe code, or library functions which used unsafe blocks internally, but exposed a safe interface (a programming pattern known as ``interior unsafe'').
In larger codebases, \cite{qin2020understanding} listed numerous cases of memory bugs related to unsafe code, but they found that this has more to do with the complexity of the type system (and lifetimes especially), and not developer negligence.
And good programming patterns such as interior unsafe, best-practices such as coding reviews, and better tools for reasoning about lifetimes and unsafe code could alleviate many causes for these bugs.

So while Rust is not as popular as comparable languages which lack its advanced type system, developers do like it and try to adopt it, mainly for its promise of safety.
Though alongside safety, they also value its performance, lack of a garbage collector, and great tooling.
And as much as they appreciate Rust's safety, they still have numerous justified reasons for using unsafe code, and do so responsibly.

\section{Future Work}\label{sec:future}
In this section, I will briefly introduce ideas from the fields of HCI and Computer Science Education that I believe could contribute to better understanding and improving the usability of Rust, or indeed any other advanced type system.
These ideas are eclectic, and I mean to introduce them as a starting point for generating ideas, not as fully fleshed-out research plans.
Also, as the focus of this report is on human-centered approaches, I will not discuss techniques from Programming Languages and Compilers research.
Though, especially if combined with HCI techniques, they could be essential to improving the usability of Rust as well.

\subsection{Better Error Messages}\label{subsec:errormsg}
Perhaps the most immediately actionable takeaway from \autoref{sec:usability} is that Rust's error messages are an important limitation.
But, paradoxically, the general community consensus is that Rust's error messages are more helpful than most languages \cite{fulton2021benefits}.
So a good next step in improving Rust's usability, and an important consideration when designing any language with such complex types, is to see where Rust's error messages succeed and how they fail.

A great starting point for this is \cite{becker2019compiler}.
They survey all works on compiler error messages in the past 50 years, and provide a number of remarkable insights on the subject.
For one, they discuss empirical evidence showing that programmers, both novice and expert, \emph{do} read compiler error messages \cite{barik2017do,prather2017on}, and so improving error reporting is indeed worthwhile.

They also compiled a list of empirically-backed guidelines for designing error messages, which can be invaluable as a shared foundation for synthesizing various works on Rust's usability.
For example, it could serve as the complementary theoretical background to empirical analyses (such as in \cite{zhu2022learning}) that argue that Rust's error messages are notably well-designed\footnote{A deeper look into this is outside the scope of this report, but my personal experience suggests that Rust follows most of the guidelines from \cite{becker2019compiler} most of the time.}.

It is also a good resource for understanding why Rust error messages fail, and how to improve them.
For instance, one of the important guidelines for error design is to include the relevant context of the error directly in the message, which multiple works have argued Rust sometimes fails to do \cite{dominik2018thesis,blaser2019thesis,zhu2022learning}.

Following \cite{bronzegc}'s finding that Rust diagnostics lack necessary architectural hints, the guidelines would also be invaluable in providing the human-centered design element to programming languages or machine learning techniques that could identify such architectural changes and present them to the programmer.

\subsection{Program Visualization Tools}

\begin{figure}
    \centering
    \includegraphics[width=0.75\textwidth]{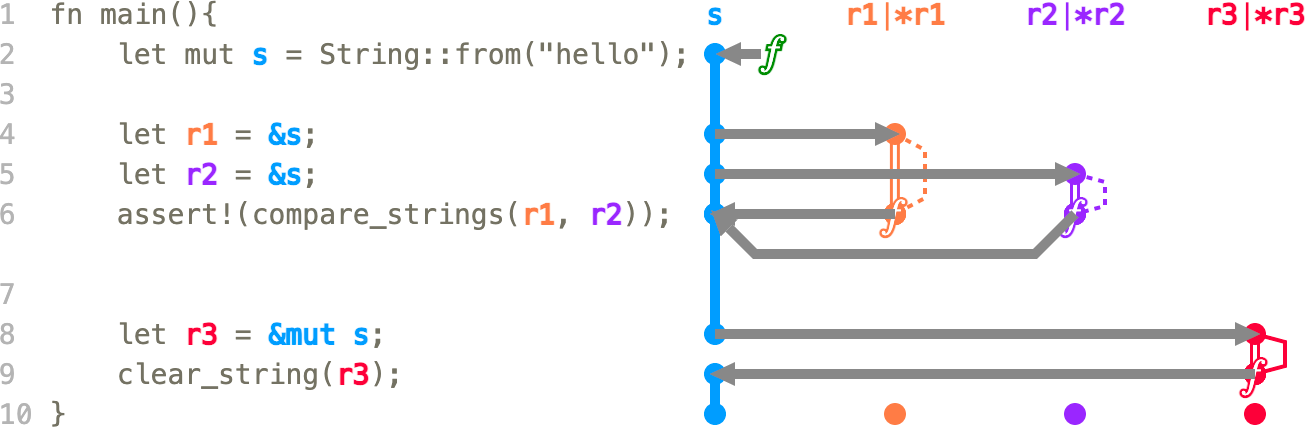}
    \caption{A RustViz visualization. If the user hovers over the nodes and arrows on the right, it displays a pop-over containing additional information about the event.}
    \label{fig:rustviz}
\end{figure}

An often mentioned next step for the usability of Rust's Ownership is visualization of Rust's lifetimes.
In their discussion of ways to address the usability issues with the incompleteness of the borrow-checker, \cite{crichton2020usability} recommended further research into visualizing the various static information provided by the borrow-checker, citing the difficulty in visualizing the large amount of information in a ``succinct, non-intrusive, yet informative'' manner.

Similarly, \cite{qin2020understanding} suggested an IDE plugin for visualizing lock lifetimes.
They found that a notably common cause of deadlock bugs in multi-threaded code was that in Rust unlocking a \rcode{Mutex} occurs implicitly when the locked value is dropped.
To address this, they recommended ``plug-ins to highlight the location of Rust's implicit unlock'',
which is a specialized case of visualizing lifetimes.

There are three existing but limited works attempting to visualize Ownership and lifetimes in Rust.
The first is RustViz \cite{gongming2020rustviz}, a tool for building visualizations of Ownership and borrowing events in Rust programs.
RustViz does not generate these visualizations automatically, but instead requires (arguably considerable) effort on the part of a ``teacher'' to specify them by annotating the code being visualized, and using a Rust library to generate it.
The benefit is that the generated visualization is interactive: the ``learner'' can hover over parts of the image to get a brief description of the events.
See \autoref{fig:rustviz} for an example of RustViz's visualizations.

The other two are related Bachelor's Theses \cite{dominik2018thesis,blaser2019thesis}.
In the first \cite{dominik2018thesis}, the author uses Polonius (an experimental implementation of the borrow-checking rules in Datalog) to extract lifetime constraints from a given program, and visualize it as a directed graph.
They motivated it by arguing that \rustc error messages do not always present all the necessary information for Ownership errors, and that this visualization could address that problem.

\begin{figure}
    \centering
    \begin{subfigure}[c]{0.5\textwidth}
        \begin{minted}[fontsize=\footnotesize,linenos,autogobble]{rust}
            fn main() {
                let mut x = 4;
                let y = foo(&x);
                let z = bar(&y);
                let w = foobar(&z);
                // ...
                x = 5;
                take(w);
            }

            fn foo<T>(p: T) -> T { p }
            fn bar<T>(p: T) -> T { p }
            fn foobar<T>(p: T) -> T { p }
            fn take<T>(p: T) { unimplemented!() }
        \end{minted}
    \end{subfigure}
    \begin{subfigure}[c]{0.25\textwidth}
        \includegraphics[width=1\textwidth]{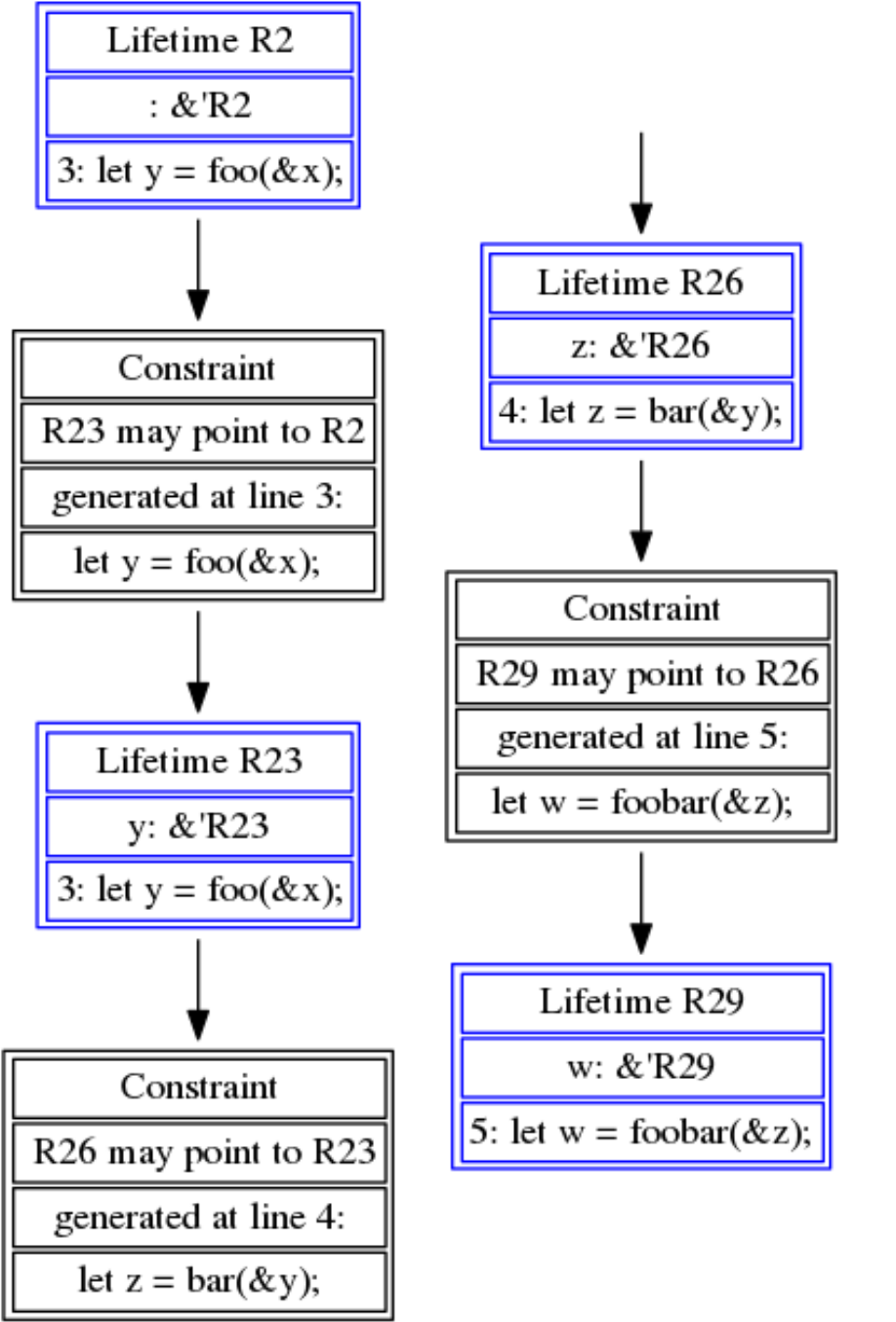}
    \end{subfigure}
    \caption{A Rust Life Assistant visualization, showing the lifetimes of the references in the code, and the corresponding constraints.}
    \label{fig:lifeassistant}
\end{figure}

\cite{blaser2019thesis} then extended \cite{dominik2018thesis} to improve its usability.
They defined an algorithm which filters the complete graph to a single path that contains all constraints relevant to the particular error message, such as in \autoref{fig:lifeassistant}.
They also created a Visual Studio Code extension called ``Rust Life Assistant'' for displaying these graphs in the IDE, and an algorithm for converting the graph into a bullet-point list of English text explaining the cause of the error.

All these works are quite limited, and crucially none of them have been evaluated on users\footnote{\cite{gongming2020rustviz} contains a study proposal, but as of writing this report, no results of such a study have been published.}.
So a great next step could be to design, implement and evaluate a tool for visualizing Ownership and lifetimes.
Such a tool could be pedagogical (similar to RustViz) or utilitarian (similar to Rust Life Assistant), but as \cite{zhu2022learning} points out ``learning Rust is a continuous process'', and I would recommend considering existing works on education and program visualization systems to inform the design and evaluation of any such tool.

A great survey of such systems can be found in \cite{sorva2013review}.
Besides exploring over 40 program visualization tools, they also provide a taxonomy of visualization systems, and emphasize that \emph{how users engage} with a visualization tool is as important as the design of the tool itself.
Specifically, they cite \cite{hundhausen2002meta} who performed a meta-study of 24 algorithm visualization (AV) tools and found that ``the form of the learning exercise in which AV technology is used is actually more important than the quality of the visualizations produced by AV technologies.''
Unfortunately most tools focus on the runtime behavior and values of programs, but they can still help us better think about the design of a visualization system, and users' engagement with it.

\subsection{Grounded Theory}\label{subsec:groundedtheory}
The works in \autoref{sec:usability} use a wide variety of research methods, including surveying existing experience reports, semi-structured interviews, online surveys, and controlled experiments.
But so far they have left a notable gap in the methodologies, which is a deep qualitative understanding of how Rust programmers actually write code.
Questions such as what tools they use, if and how they read error messages, how they reason about various errors, common debugging strategies, etc. are all left out of the scope of the existing research.
To fill this gap, I recommend building a Grounded Theory of how Rust programmers write code in relation to programmers in other languages, or using different type or memory management systems.

Grounded Theory (GT) started in the 1960s as a research method in sociology, but has since become a standard method of qualitative research in many field as a method for developing theories bottom-up through observation \cite{charmaz2010grounded}.
Speaking very broadly, GT as a method is an iterative process of collecting data through interviews and observations and using open-coding to develop a theory that is \emph{grounded} in the empirical data (rather than informed by or confirming existing theories).
But the above description is too simplistic.
Various versions of GT exist, each of which make different assumptions about the nature of knowledge (positivism vs. constructivism) and the precise steps they follow are subtly different and outside the scope of this report.

That said, GT is an established and popular method in Software Engineering research \cite{stol2016grounded},
and recently \cite{lubin2021how} employed \emph{Constructivist} GT \cite{charmaz2006constructing} to develop a deeper understanding of how statically-typed functional programmers write code.
Which is why I recommend building a grounded theory of ``How Rust Programmers Write Code''.
Speculating on what we may learn from this is counter to GT's philosophy.
But my hope is that such a GT will help develop the foundations that can both better motivate and help us understand the human-centered study and design choices involved in Rust and similar advanced type systems.

%% Acknowledgments
\begin{acks}
    I would like to thank my advisors Sorin Lerner and Nadia Polikarpova for providing feedback on this report, as well as their advice and mentorship. I would also like to thank Hila Peleg for being my de-facto advisor in my first two years in Graduate school, and Ruanqianqian (Lisa) Huang and Shraddha Barke for being amazing collaborators and co-authors whose research has been a major source of motivation and inspiration in my own work.
\end{acks}
\bibliography{citations}

%% Appendix
% \appendix
% \section{Appendix}

\end{document}